 \documentclass[manuscript]{aastex62}

\received{Nov. 24, 2019}
\revised{Jan. 10, 2020}
\accepted{Jan. 15, 2020}
\submitjournal{AJ}

\shorttitle{Resonances in non-axisymmetric potentials}
\shortauthors{Sicardy}

\begin{document}

\title{Resonances in non-axisymmetric gravitational potentials}

\correspondingauthor{Bruno Sicardy}
\email{bruno.sicardy@obspm.fr}

\author[ 0000-0003-1995-0842]{Bruno Sicardy}
\affil{LESIA, Observatoire de Paris, \\
Sorbonne Universit\'e, Universit\'e PSL, CNRS, \\
Univ. Paris Diderot, Sorbonne Paris Cit\'e, \\
5 place Jules Janssen, 92195 Meudon, France}



\begin{abstract}
We study sectoral resonances of the form $j\kappa= m(n-\Omega)$
around a non-axisymmetric body with spin rate $\Omega$, where 
$\kappa$ and $n$ are the epicyclic frequency and mean motion of a particle, respectively,
where $j>0$ and $m$ ($<0$ or $>0$) are integers, $j$ being the resonance order.
This describes $n/\Omega \sim m/(m-j)$ resonances inside and outside the corotation radius, 
as well as prograde and retrograde resonances.
Results are: 
(1) the kinematics of a periodic orbit depends only on $(m',j')$, 
the irreducible (relatively prime) version of $(m,j)$.
In a rotating frame, 
the periodic orbit has $j'$ braids, $|m'|$ identical sectors and $|m'|(j'-1)$ self-crossing points;
(2) thus, Lindblad resonances (with $j=1$) are free of self-crossing points;
(3) resonances with same $j'$ and opposite $m'$ have the same kinematics,
and are called \textit{twins};
(4) the order of a resonance at a given $n/\Omega$ depends on the symmetry of the potential.
A potential that is invariant under a $2\pi/k$-rotation creates only resonances with $m$ multiple of $k$;
(5) resonances with same $j$ and opposite $m$ have the same kinematics and same dynamics, 
and are called \textit{true twins};
(6) A retrograde resonance ($n/\Omega < 0$) is always of higher order
than its prograde counterpart ($n/\Omega > 0$);
(7) the resonance strengths can be calculated in a compact form with the classical operators
used in the case of a perturbing satellite.
%
%
Applications to Chariklo and Haumea are made.
\end{abstract}



\section{Introduction} \label{sec_intro}

Resonances between a non-axisymmetric rotating potential and orbiting particles 
have a very vast domain of applications, 
from galactic disks perturbed by a central bar to spiral waves excited in Saturn's rings by satellites
or planetary modes. 
More recent examples are given by dense rings discovered around the small
Centaur object Chariklo in 2013 \citep{bra14}, 
and the trans-neptunian dwarf planet Haumea in 2017 \citep{ort17}.
%
%
%
Both objects significantly depart from axisymmetric shapes, 
and are thus expected to drive strong resonances in their respective circum-body collisional disks \citep{sic19,sic20}.

For a better understanding of these disks, it is important 
to clarify and classify the kinematics and dynamics of the various resonant orbits.
Also, considering the variety of shapes assumed by those bodies,  
a simple numerical scheme to calculate the resonance strengths is desirable.
In the case of a perturbing satellite, resonance strengths are classically calculated
by using operators (denoted $F_n$ in this paper) acting on Laplace coefficients. 
Those operators can be found in various publications,
see for instance \cite{mur00} and \cite{ell00}, MD00/EM00 herein.
Here I show that the $F_n$ operators can in fact be used formally for {\it any} non-axisymmetric potential, 
provided that some symmetry conditions are met.
Those operators encapsulate in a single expression 
the direct and indirect terms of the potential, 
as well as  inner and outer resonances
(lying inside and outside the corotation radius, respectively),
and and account for both prograde or retrograde particle motions.

This paper is organized as follows: 
Section~\ref{sec_prelem} provides the general context of the study,
Section~\ref{sec_reson} classifies the various resonances that occur in that context,
Section~\ref{sec_expan} describes the structure of the resonant orbits (kinematics and dynamics),
Section~\ref{sec_strength} shows how the $F_n$ operators mentioned above can be used for a generic potential,
with applications to Chariklo and Haumea, assumed to be homogeneous triaxial ellipsoids.
Section~\ref{sec_concl} provides concluding remarks.

\section{Preliminay remarks} 
\label{sec_prelem}

We consider a body rotating at constant angular speed $\Omega= 2\pi/T_{\rm rot}$,
where $T_{\rm rot}$ is the rotation period.
The following simplifying assumptions are made: 

$(i)$ The body rotates around one if its principal axes of inertia, {\it i.e.} without wobbling motion; 

$(ii)$ The mass distribution of the body is symmetrical with respect to a plane perpendicular to the rotation axis, 
called the equatorial plane; 

$(iii)$ The mass distribution possesses a plane of symmetry that contains the rotation axis. 

An example is a spherical object with a mass anomaly sitting at its equator.
%
Another example is a triaxial homogeneous ellipsoid rotating around its smallest axis.
Further examples are given by sectoral resonances stemming for normal modes in a gaseous 
planet\footnote{This is a particular case of tesseral resonances, where the potential depends only on longitude, not latitude}.

The time-averaged gravitational potential created by the body is axisymmetric. 
From hypothesis $(i)$, 
the average vertical angular momentum (the component that is parallel to the rotation axis) of the orbiting particles is conserved.
Consequently, a dissipative collisional set of particles surrounding the body settles into the equatorial plane, 
as this configuration minimizes energy for a constant vertical angular momentum.
From hypothesis $(ii)$, no vertical forces are exerted on the equatorial disk,
so that no vertical resonances will be considered here. 

Notations are classical: the position vector \textbf{r} of a particle in the equatorial plane (counted from the center of mass of the body)
is expressed in polar coordinates $(r,L)$, 
where $r=||{\bf r}||$ and $L$ is the true longitude counted from an arbitrary origin.
The orientation of the body is measured by the longitude
$\lambda'=  \Omega t$ of a reference point on its equator,
$t$ being the time.
The motion of the particle is described by its keplerian orbital elements $a,e,\lambda,\varpi$, 
{\it i.e.} the semi-major axis, orbital eccentricity, mean longitude and longitude of pericenter, respectively.

In an inertial frame, a particle is submitted to a time-dependent potential. 
From hypotheses $(i)$ and $(ii)$, this potential takes the form $U(r,\theta)$,
where $\theta= L-\lambda'= L - \Omega t$.
This time-dependence is eliminated by writing the equations of motion in the frame co-rotating with the body.
In that frame, the energy of the particle is a constant of motion, called the Jacobi constant.
As it moves in the equatorial plane, the particle has two degrees of freedom, 
each associated with a fundamental frequency. 
One is the radial epicyclic frequency $\kappa= n - \dot{\varpi}$ (where the dot denotes time derivative),
the frequency at which the particle returns to its pericenter, and
the other is the synodic frequency $n-\Omega$, the frequency at which the particle returns to a fixed 
position relative to the body. 

As $U(r,\theta)$ is $2\pi$-periodic in $\theta$, 
it can be Fourier-expanded as $U({\bf r})= \sum_{m=0}^{+\infty} U_m(r) \cos(m\theta+\varphi_m)$,
where $U_m$ and $\varphi_m$ are uniquely defined.
From hypothesis $(iii)$,
$U(r,\theta)$ is an even function of $\theta$ if the reference point on the body 
is taken in the vertical plane of symmetry. Then $\varphi_m=0$ and
\begin{equation}
U({\bf r})= \sum_{m=0}^{+\infty} U_m(r) \cos(m\theta).
\label{eq_U0}
\end{equation}
Here, the integer $m$ is called the \textit{azimuthal number}.
It describes the number of cycles completed by the potential during one revolution around the body.
A more symmetric form can be adopted, in which $m$ assumes both positive and negative values, 
\begin{equation}
U({\bf r})= \sum_{m=-\infty}^{+\infty} U_m(r) \cos(m\theta),~{\rm with}~U_{(-m)}=U_{(m)},
\label{eq_U}
\end{equation}
the parity condition ensuring the unicity of the coefficients $U_m(r)$.
In this case, each coefficient $U_m(r)$ is divided by two compared to its value in Eq.~\ref{eq_U} 
(except for $U_0(r)$, which remains unchanged).
Choosing between Eqs.~\ref{eq_U0} and \ref{eq_U} is arbitrary.
Here I choose Eq.~\ref{eq_U} as it offers a more natural way to expand the potential 
in resonant terms, see Section~\ref{sec_expan}.

\section{Resonance taxonomy} 
\label{sec_reson}

The potential $U(r,\theta)$ can be Fourier-expanded along linear combinations of the fundamental frequencies
$\kappa$ and $n-\Omega$, 
\begin{equation}
\nu_{j,m} = j\kappa - m(n-\Omega),
\label{eq_nu_mj}
\end{equation}
where $m$ can be positive or negative, see above.
Without loss of generality, the integer $j$ can be taken as always positive.
It is the order of the resonance, see sub-Section~\ref{subsec_order}.
Resonances occur for $\nu_{j,m} = 0$.
If $j=0$ then
\begin{equation}
n= \Omega,
\end{equation}
called the corotation resonance.
This resonance is discussed in the context of elongated bodies 
by \cite{sch94,sic19} and will not be consider further here.
%
%
Thus, we restrict ourselves to the case $j > 0$, and the resonance condition reads
\begin{equation}
j\kappa = m(n-\Omega).
\label{eq_reson}
\end{equation}
It means that after $|m|$ radial oscillations, 
the particle completes exactly $j$ synodic period around the body.
This excites the orbital eccentricity of the particles, a way to create a coupling between the disk and the body.
%
%
From $\dot{\varpi}= n - \kappa$, Eq.~\ref{eq_reson} can be re-expressed as
\begin{equation}
\frac{n-\dot{\varpi}}{\Omega-\dot{\varpi}} = \frac{m}{m-j},
\label{eq_ratio_n_omega_exact}
\end{equation}
meaning that in a frame rotating at the particle precession rate $\dot{\varpi}$, 
the particle completes $m$ revolutions while the body completes $m-j$ rotations,
hence the notation ``$m/(m-j)$" 
resonance\footnote{In galactic dynamics, $\kappa$ and $\Omega$ 
are usually very different so that this approximation does not hold, and the notation $m/(m-j)$ resonance is 
meaningless. Instead, the cases corresponding to $j=1$ are sometimes referred to as a $m:1$ (Lindblad) resonance,
see {\it e.g.} \cite{pfe84}.}.
The case $m=j$ corresponds to the apsidal resonance $\Omega = \dot{\varpi}$,
in which the particle's orbit precesses at the rotation rate of the body.
For moderately non-axisymmetric potentials, we have $\dot{\varpi} \ll \Omega$, so that apsidal resonances do not occur
and this case is not studied here. 
Eq.~\ref{eq_ratio_n_omega_exact} can then be written
\begin{equation}
\frac{n}{\Omega} \sim \frac{m}{m-j}.
\label{eq_ratio_n_omega_approx}
\end{equation}
%

\subsection{Location of resonances}

The axisymmetric part of the potential (the term $U_0(r)$ in Eq.~\ref{eq_U})
provides $n$ and $\kappa$ \citep{cha42}: 
\begin{equation}
\begin{array}{ccc}
\displaystyle
n^2(r)= \frac{1}{r} \frac{dU_0(r)}{dr} &
{\rm and} & 
\displaystyle
\kappa^2(r)= \frac{1}{r^3} \frac{d(r^4 n^2)}{dr}. \\
\end{array}
\label{eq_n_kappa}
\end{equation}

The condition $j\kappa = m(n-\Omega)$ then allows the calculation of the resonance location, 
see a practical example in Section~\ref{sec_strength}.

\subsection{Prograde and retrograde resonances}

A debris disk around a body may result from an impact, 
so that rings may move in two opposite directions, 
prograde or retrograde\footnote{Retrograde motions may also be encountered
with exo-planets orbiting a circular binary stellar system, see \cite{mor12}.}.
Retrograde resonances then occur for $n/\Omega \sim m/(m-j) < 0$.
As $j>0$, this occurs for 
\begin{equation}
0 < m < j,
\label{eq_retro_reson}
\end{equation}
while prograde resonances occur for
\begin{equation}
m< 0 {\rm~or~} j < m.
\label{eq_progr_reson}
\end{equation}

Note in passing that in Eq.~\ref{eq_reson}, $\kappa$ and $n$ must have the same sign
in order to consistently describe a progade or retrograde motion. 
Adopting arbitrarily $\Omega >0$,
a prograde orbit has $\kappa,n>0$, while a retrograde orbit has $\kappa,n<0$.

%

\subsection{Inner and outer resonances}

The position of a resonance can be interior to the corotation radius (where $n= \Omega$),
in which case we talk about an inner (or internal) resonance, and $|n/\Omega| > 1$.
If the resonance occurs outside the corotation radius, 
we talk about an outer (or external) resonance, and $|n/\Omega| < 1$.
From Eq.~\ref{eq_ratio_n_omega_approx}, resonances with $m<0$ are always external.
For $0 < j < 2m$, the resonances are internal and for $2m < j$, they are external.
The case $j=2m$ corresponds to the ``retrograde corotation resonance", 
in which the particle moves at the corotation radius, but opposite to the 
body\footnote{%
The term retrograde corotation is in fact not appropriate because 
the particle mean motion does not match any harmonics of the potential. 
Actually, this resonance has the same dynamical behavior as the prograde 1/3 resonance, 
see sub-Section~\ref{subsec_struct}.
However, to keep in line with the nomenclature of other publications, 
I still use in the terms ``retrograde corotation" in the text.}.

\subsection{Lindblad resonances}

Here, I restrict the term \it Lindblad resonances \rm to first-order resonances ($j=1$). 
In the literature, Eq.~\ref{eq_reson} is usually written as $\kappa = \pm m(n-\Omega)$, with the convention $m>0$.
This introduces the presence of numerous $\pm$ and $\mp$ symbols in the equations,
a possible source of errors. 
In contrast, taking both positive and negative values for $m$
eases the calculations by avoiding the cumbersome use of the $\pm$ symbol.

\begin{figure}
\plotone{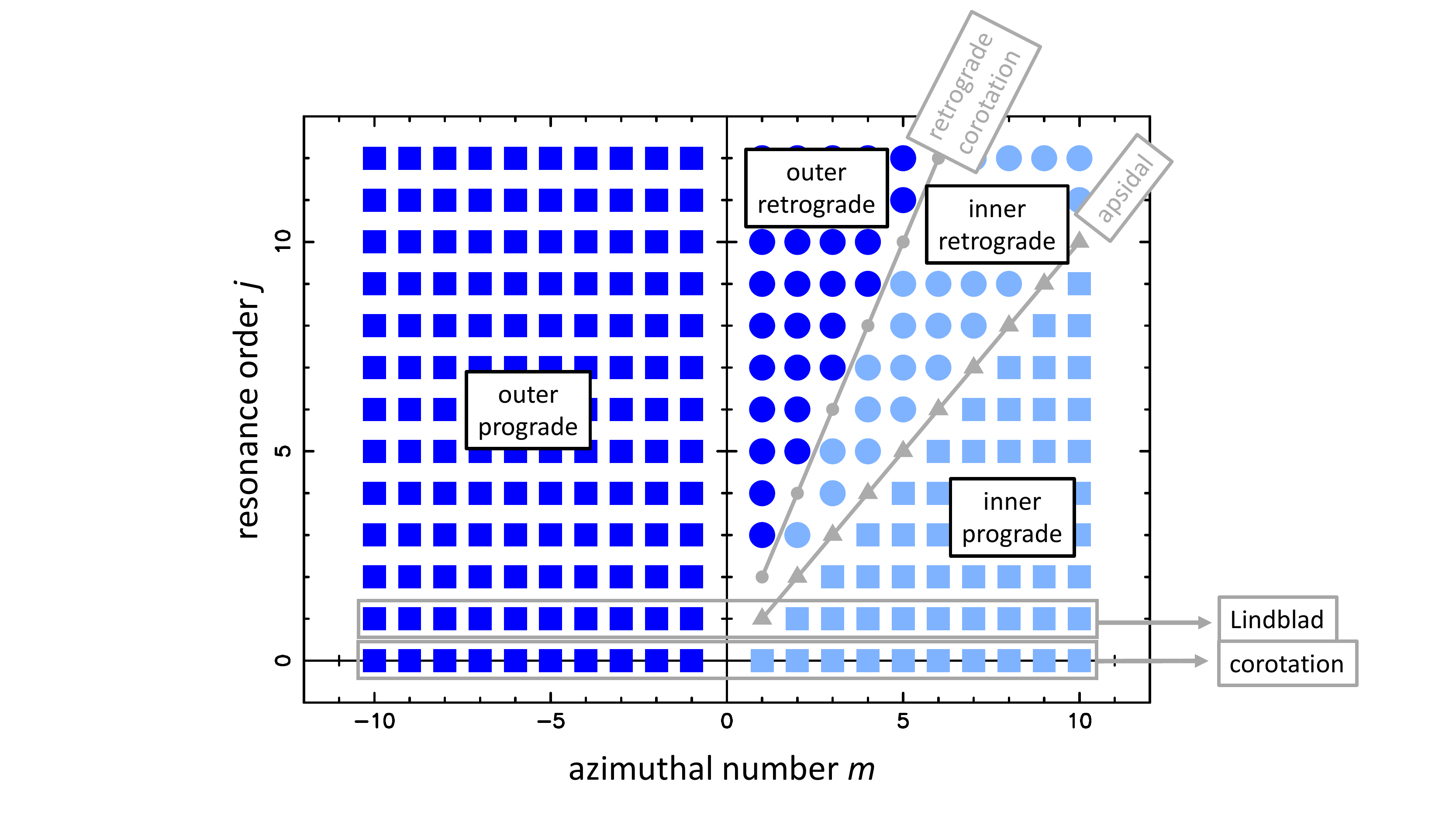}
\caption{%
Resonance taxonomy in a $(m,j)$ diagram, 
where $m$ is the azimuthal number and $j$ is the resonance order.
Squares (resp. dots) are for prograde (resp. retrograde) resonances.
Darker (resp. lighter) blue is for outer (resp. inner) resonances
Gray triangles are for apsidal resonances 
(not treated here) 
and dots are for the retrograde corotation. 
}%
\label{fig_taxonomy}
\end{figure}

The various resonances described in this Section are summarized in Fig.~\ref{fig_taxonomy} in a $(m,j)$ diagram.

\section{Structure of resonant orbits} 
\label{sec_expan}

The potential $U({\bf r})$ of Eq.~\ref{eq_U} can be expressed in terms of 
$\lambda'$ and $(a,e,\lambda,\varpi)$, 
and then expanded in powers of $e$ under the forms
$r/a= 1 + \sum_{j=1}^{+\infty} e^j E_j\cos^j (\lambda-\varpi)$ and
$L= \lambda + \sum_{j=1}^{+\infty} e^j L_j\sin^j (\lambda-\varpi)$,
where $E_j$ and $L_j$ are numerical coefficients that describe the keplerian motion. 
In doing so, each term $\cos(m\theta)$ in Eq.~\ref{eq_U},
when combined to the terms $\cos^j, \sin^j (\lambda-\varpi)$,
provides two terms of the form 
$e^j \cos[m \lambda' - (m-j)\lambda - j\varpi]$ and 
$e^j \cos[m \lambda' - (m+j)\lambda + j\varpi]$.
Noting that the second term can be written $e^j \cos[(-m)\lambda' - (-m-j)\lambda - j\varpi]$,
the expansion of $U({\bf r})$ may be written using only terms of the form
$m \lambda' - (m-j)\lambda - j\varpi$, where $m$ is positive or negative.
After re-ordering those terms, we obtain
\begin{equation}
U(\textbf{r}) = U(a,e,\lambda,\varpi,\lambda')= 
\sum_{k=-\infty}^{+\infty}  U_k(a) \cos\left[k(\lambda-\lambda')\right] +  
\sum_{m=-\infty}^{+\infty} \sum_{j=1}^{+\infty}  
\bar{U}_{m,j}(\alpha) e^j \cos\left[m \lambda' - (m-j)\lambda - j\varpi\right],
\label{eq_pot_expan}
\end{equation}
where
\begin{equation}
\alpha= \frac{a}{R}, 
\label{eq_alpha}
\end{equation}
$R$ being a characteristic dimension of the problem (to be defined later).
Note that the first summation in Eq.~\ref{eq_pot_expan} describes the corotation resonance.

A given $m/(m-j)$ resonance occurs for $\nu_{m,j}=0$, {\it i.e.} for $m \lambda' - (m-j)\lambda - j\varpi$ stationary.
Let us denote $U_{m,j}$ the potential associated with that resonance, {\it i.e.} 
\begin{equation}
U_{m,j}(a,e,\lambda,\varpi,\lambda')= \bar{U}_{m,j}(\alpha) e^j \cos\left(j\phi_{m,j}\right),
\label{eq_Umj_def}
\end{equation}
where
\begin{equation}
\phi_{m,j}= \frac{m \lambda' - (m-j)\lambda - j\varpi}{j}
\label{eq_phi_mj}
\end{equation}
is the resonant argument.
Note the dividing factor $j$, which ensures that the proper choice of canonical variables is made
for use in the Hamiltonian describing the resonance \citep{pea86}.

\subsection{Resonance order}
\label{subsec_order}

The term $U_{m,j}$ is of order $e^{j}$, 
a property known as the d'Alembert's characteristics, and $j$ is called the order of the 
resonance\footnote{Higher order terms in eccentricity are in fact present in the amplitude of 
the term $\cos(j\phi_{m,j})$, but there are ignored here.}.
The order $j$ is not entirely determined by the ratio $n/\Omega$, 
as the same value $n/\Omega = m/(m-j)$ can be achieved with
multiples of $m$ and $j$.
Let us denote $(m',j')$ the relatively prime (or irreducible) version of $(m,j)$.
Then the same ratio $n/\Omega$ is achieved for all couples of the form $(km',kj')$, $k$ integer.
Thus, at the same radius, an infinity of resonances of orders $j'$, $2j'$,... exist.
Usually, only the resonance of lowest order, $j'$, is considered, 
and the higher-order, weaker resonances are neglected.

The symmetry of the potential, however, may lead to the vanishing of some resonances.
If the potential is invariant under a rotation of $2\pi/k$ radians 
(as it is the case for normal sectoral modes in gaseous planets, for instance), 
then only $m$'s that are multiples of $k$ appear in Eq.~\ref{eq_U}. 
Thus, the ratio $n/\Omega$ takes the form
\begin{equation}
\frac{n}{\Omega} \sim  \frac{m}{m-j} = \frac{kp}{kp-j}.
\label{eq_ratio_n_omega_approx_k}
\end{equation}

Consequently, only $m/(m-j)$ resonances where $m$ is multiple of $k$ survive in a $2\pi/k$-periodic potential.
For instance, every other $k$ Lindblad resonances ($j=1$) remain in this context.
This is discussed in Section~\ref{subsec_homo_ellipsoid} with the potential of a triaxial ellipsoid,
which is invariant under a $\pi$-rotation ($k=2$).
Then, only every other Lindblad resonances survives, 
those with even values of $m$, see Eq.~\ref{eq_ratio_n_omega_approx_ell}.
Similarly, the second-order 1/3 resonance vanishes, leaving the fourth-order 2/6 resonances at its place.
The distinction is important because, although corresponding to the same ratio $n/\Omega$, 
these two resonances have different phase portraits and different dynamical behaviors.
To make that distinction clear, the 2/6 notation should not be simplified to 1/3.

\subsection{Structure of the periodic resonant orbits}
\label{subsec_struct}

\begin{figure}[!t]
\plotone{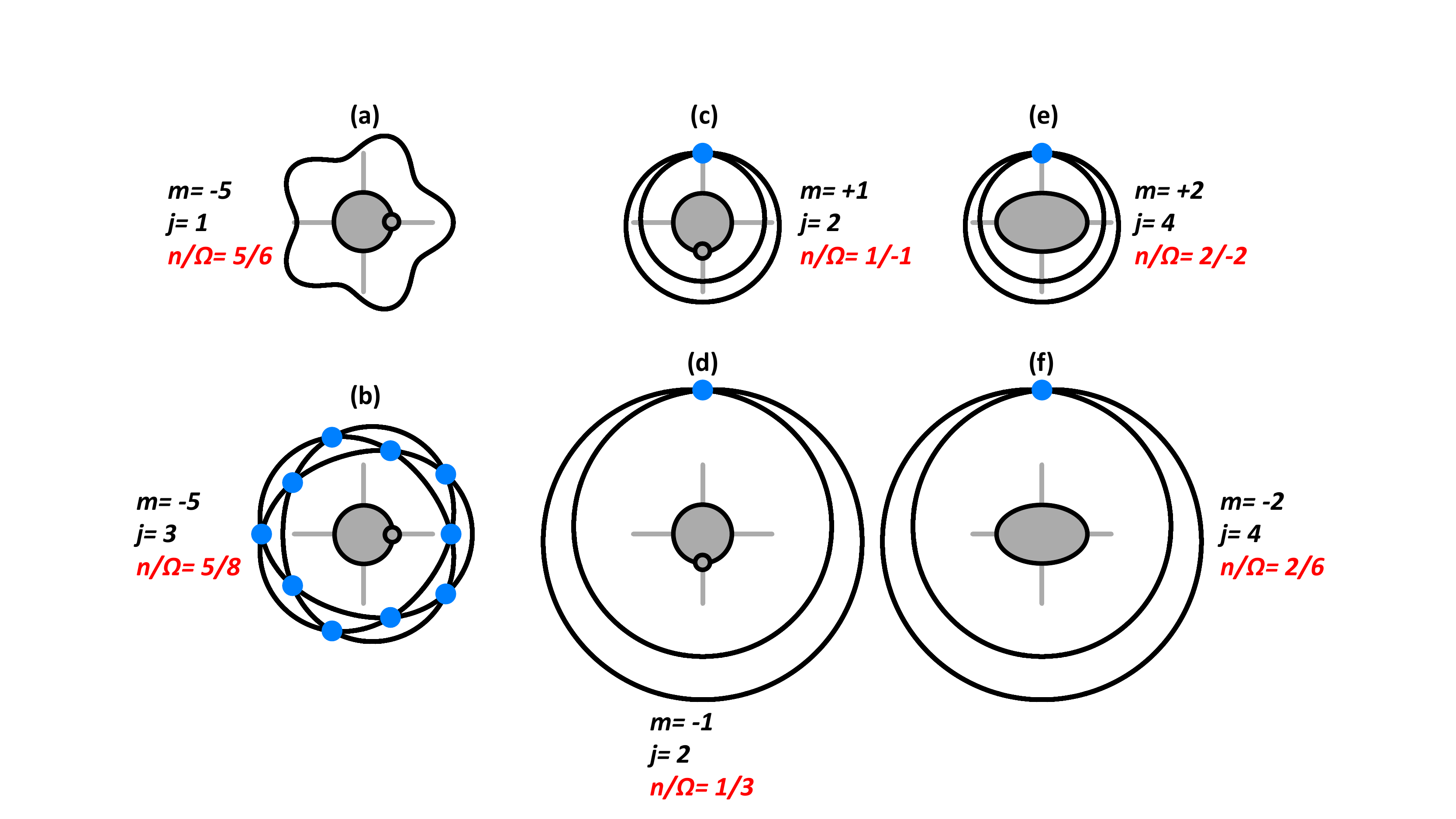}
\caption{%
Pole-on view of various $m/(m-j)$ resonant periodic orbits around a body 
that is either a sphere with a mass anomaly sitting at its equator, 
or an elongated ellipsoidal object. 
The grey crosses correspond to the radius of the corotation orbit, 
where particles revolve at the same angular speed as the spin rate of the body.
Each orbit has an eccentricity of 0.15. 
The blue dots mark the self-crossing points.
(a) The periodic orbit corresponding to the 5/6 (first-order) outer Lindblad resonance.
(b) The same for the 5/8 outer (third-order) resonance.
The orbit has 3 braids, 5 identical sectors and $|m|(j-1)=10$ self-crossing points,
thus satisfying Eq.~\ref{eq_Nc}.
(c) The retrograde ``corotation" orbit, 
actually corresponding to the second-order retrograde resonance 1/-1.
(d) The true twin of Case (c), corresponding to the 1/3 outer resonance.
The two orbits (c) and (d) have the same kinematics and same dynamical behaviors. 
They also are the only periodic orbits with a single self-crossing point ($|m'|(j'-1)=1$).
(e) The same as Case (c), but with an ellipsoidal central body.
The resonance is now of order four and is labelled as 2/-2.
The cases (c) and (e) correspond to false twins, with same kinematics but different orders,  
hence different dynamical behaviors. 
(f) The true twin orbit of Case (e), corresponding to the 2/6 fourth-order outer resonance.
}%
\label{fig_braids_self_crossing}
\end{figure}

Contrarily to the order, 
the kinematic structure of a resonant orbit depends only on the ratio $n/\Omega$,
independently of the symmetry of the potential.
The polar equation (in a frame rotating with the body) of a $m/(m-j)$ resonant periodic orbit is
\begin{equation}
\rho (\theta)= 
a \left[ 1 - e \cos \left(\frac{m}{j} \theta + \phi_{m,j} \right) \right].
\label{eq_streamline}
\end{equation}
It can also be viewed as the polar equation of a perturbed streamline, 
where the particles move at different longitudes while sharing a common $\phi_{m,j}$.
This aspect in discussed in Section~\ref{sec_concl}.
The structure of the periodic orbits is studied in details in \cite{sic20}.
Noting again $(m',j')$ the irreducible version of $(m,j)$, results are (see also Fig.~\ref{fig_braids_self_crossing}):

\begin{enumerate}
\item the periodic orbit has $j'$ distinct braids,
\item the periodic orbit is invariant by a rotation of $2\pi/|m'|$, \textit{i.e.} it possesses $|m'|$ identical sectors,
\item in each sector, the periodic orbit has $(j'-1)$ self-crossing points,
\item thus, the total number of self-crossing points 
is\footnote{The eccentricity $e$ must be small enough to obtain only the essential self-crossing points, 
that are present even for vanishingly small eccentricities.}:
\begin{equation}
N_c = |m'|(j'-1),
\label{eq_Nc}
\end{equation}
\item consequently, Lindblad resonances ($j=1$) do not lead to self-crossing\footnote{The reciprocal is not true. 
For instance the second-order 2/4 resonant orbit has no self-crossing point,
but it is not a Lindblad resonance.},
\item
the only resonances that result in a unique self-crossing point are
for $|m'|=1$ and $j'=2$, corresponding to the second-order prograde 1/3 and retrograde 1/-1 resonances. 
\item the lowest possible order of a retrograde resonance is $j'=2$ (obtained with $m'=1$ and $n/\Omega=$ 1/-1). 
Thus, there are no retrograde Lindblad resonances, 
\item resonances having the same $|m'|$ and the same $j'$
correspond to orbits that have the same shape, and more precisely, that are homothetic.
They have the same number of sectors, braids and self-crossing points,
\textit{i.e.} they have the same kinematic behavior.
I qualify two such resonances as \textit{twins}.
Note that twin resonances correspond to different ratios $|n/\Omega|$.
\item two resonances that have the same $|m|$ and $j$ have not only the same kinematic behavior,
but also the same order, {\it i.e.} the same dynamical behavior. 
I qualify two such resonances as \textit{true twins}.
(while \textit{false twins} are two resonances that have the same $|m'|$ and $j'$, but different $|m|$ and $j$).
An example of true twins are the 1/3 and 1/-1 resonances mentioned 
above\footnote{An example of a 1/-1 resonance is the retrograde asteroid 2015 BZ$_509$
that shares Jupiter's orbit \citep{wie17,mor17}.
An example of a 1/3 resonance is given by the Trans-Neptunian Object (136120) 2003 LG$_7$
that completes one prograde orbit while Neptune completes 
three (https://minorplanetcenter.net/db\_search/show\_object?object\_id=136120).}.
Fig.~\ref{fig_taxonomy} shows that
any outer prograde resonance ($m<0$) has a true twin that is either a retrograde (inner or outer) resonance, 
or an inner prograde resonance.
\end{enumerate}

The same ratio $|n/\Omega|$, and thus the same orbital radius, corresponds to two different resonances, 
one prograde ($n/\Omega>0$) and one retrograde ($n/\Omega<0$).
This is achieved for a pairs of $(m'_{\rm p},j'_{\rm p})$ and $(m'_{\rm r},j'_{\rm r})$ satisfying 
$m'_{\rm p}/(m'_{\rm p}-j'_{\rm p}) = -m'_{\rm r}/(m'_{\rm r}-j'_{\rm r})$, where the subscripts ``p" and ``r" refer 
to prograde and retrograde motions, respectively.
The couples $(m'_{\rm p},j'_{\rm p})$ and $(m'_{\rm r},j'_{\rm r})$ being each irreducible, 
so are both couples $(m'_{\rm p},m'_{\rm p} - j'_{\rm p})$ and $(m'_{\rm r},m'_{\rm r} - j'_{\rm r})$.
Gauss' theorem thus implies that $|m'_{\rm r}|=|m'_{\rm p}|$.
More precisely, since $m'_{\rm r} > 0$ (Fig.~\ref{fig_taxonomy}), we must have $m'_{\rm r}=|m'_{\rm p}|$.
Distinguishing the cases $m'_{\rm p} = m'_{\rm r}$ and  $m'_{\rm p} = -m'_{\rm r}$,
accounting for the fact that $m'_{\rm p} < 0$ or $m'_{\rm p} > j'_{\rm p}$ (Eq.~\ref{eq_progr_reson}) and
that $j'_{\rm p}, j'_{\rm r} > 0$ by convention, it is easy to show that
if $m'_{\rm p} < 0$, then $m'_{\rm r}=-m'_{\rm p}$ and $j'_{\rm r}=  j'_{\rm p} - 2m'_{\rm p}$, while
if $m'_{\rm p} > j'_{\rm p}$, then $m'_{\rm r}=m'_{\rm p}$ and $j'_{\rm r}=  2m'_{\rm p} - j'_{\rm p}$.

In all cases, $j'_{\rm r} > j'_{\rm p}$.
Thus, at a given orbital radius, retrograde resonances are always of higher order
than prograde resonances, a result already found by \cite{mor12}. 
For instance, the 3/2 prograde resonance is of order one ($m'=3, j'=1$),
while the 3/-2 retrograde resonance is of order five ($m'=3, j'=5$).

\section{Resonance strength}
\label{sec_strength}

We now calculate the terms $\bar{U}_{m,j}(\alpha)$ of Eq.~\ref{eq_pot_expan},
first in the case of a mass anomaly, 
and then generalizing the results to any potential of the form of Eq.~\ref{eq_U}.

\subsection{Mass anomaly}

We consider a spherical body of mass $M$ and radius $R$,
with a mass anomaly $m_a$ sitting on its equator.
The potential $U(\textbf{r})$ then takes the form \citep{sic19}
\begin{equation}
U(\textbf{r}) = 
\displaystyle
-\frac{GM}{r} - \frac{GM\mu}{R} 
\left\{\frac{1}{2} \sum_{m=-\infty}^{+\infty} 
\left[b_{1/2}^{(m)}\left(\frac{r}{R}\right) - q \delta_{(|m|,1)}\left(\frac{r}{R}\right) \right] 
\cos (m\theta)\right\},  
\label{eq_disturb_pot_ma}
\end{equation}
where $q= \Omega^2 R^3/GM$ is the rotation parameter, 
$\mu= m_a/M$ is the normalized mass anomaly, 
$b_{1/2}^{(m)}$ is the classical Laplace coefficient 
$b^{(m)}_{\gamma}(\alpha)= (2/\pi) \int_0^\pi \cos(m\theta)/[1+\alpha^2-2\alpha \cos(\theta)]^\gamma d\theta$
and the symbol
$\delta_{(|m|,1)}$ is the Kronecker delta function stemming from the indirect part of the potential in $U(\textbf{r})$,
while the terms $b_{1/2}^{(m)}$ describe the direct part of the potential.
This potential is formally identical to that caused by a satellite on a circular orbit (corresponding to $q=1$), 
except that the mass anomaly revolves at angular velocity $\Omega$ (instead of the keplerian velocity of a satellite)
at the surface of the body. This effect is encapsulated in the parameter $q$.

The coefficients $\bar{U}_{m,j}(\alpha)$
are calculated using the formal expansions of the disturbing potential due to a satellite on a circular orbit, 
see MD00/EM00.
For instance, consider the first-order resonant argument $\phi_{m,1}= m\lambda' - (m-1)\lambda -\varpi$ with $m>0$
corresponding to an external perturber. The aforementioned references then 
provide\footnote{Note that MD00/EM00 denote the azimuthal number $j$, while we use $m$ here.}
\begin{equation}
\bar{U}_{m,1}(\alpha) = -\frac{GM\mu}{R} f_{27},
\label{eq_f27}
\end{equation}
where
\begin{equation}
f_{27}= 
\frac{1}{2} [-2m - \alpha D][b_{1/2}^{(m)}(\alpha) - q \delta_{(|m|,1)}\alpha],~{\rm with~} D= \frac{d}{d\alpha}.
\end{equation}
The factor $f_{27}$ can be re-written
\begin{equation}
f_{27}= F_{27}\left[b_{1/2}^{(m)}(\alpha) - q \delta_{(|m|,1)}\alpha \right],
\end{equation}
where $F_{27}= (1/2) [-2m - \alpha D]$ is now a linear operator.
Note that we have included here the \it indirect part of the potential, $q \alpha \delta_{|m|,1}$. \rm
It is easy to verify that the operator $F_n$ can be generically applied to that indirect term, 
so that there is no need to look at the entries of the indirect parts in the 
tables\footnote{This point is mentioned in \cite{agn12}, p. 6.}.
Actually, since the indirect part of the potential is linear in $\alpha$,
the differential operators $D^p= d^p/d\alpha^p$ applied to the indirect term reduce the simple form
\begin{equation}
\alpha^p D^p= \alpha \delta_{(p,1)}.
\label{eq_aD}
\end{equation}
For $m<0$, then the factor usually considered in Eq.~\ref{eq_f27} is $f_{31}$ instead of $f_{27}$.
However, this complication is not necessary, as $f_{31}$ is a mere avatar of $f_{27}$, 
that can be deduced from it through the transformation $m \rightarrow 1-m$ and $\alpha \rightarrow 1/\alpha$.
Note that in doing so, we may have $\alpha > 1$.
This poses \textit{a priori} a problem from a computational point of view, 
as classical series expansions of $b_{1/2}^{(m)}(\alpha)$ use series in powers of $\alpha$
that converge only for $\alpha<1$.
However, this problem is easily resolved by using the identity 
$b_{1/2}^{(m)} (\alpha) = b_{1/2}^{(m)} (1/\alpha)/\alpha$.

The same approach can be used for any resonance $m/(m-j)$, 
considering only the entries with cosine arguments of the form
$m\lambda' - (m-j)\lambda - \varpi$ in order to find $F_n$
(in practice, the entries labeled ``4D$j$.1" in MD00/EM00).
Then the fact that the perturber is internal or external is automatically accounted for through the values of $m$ and $j$. 
So, the term $U_{m,j}(\alpha)$ in Eq.~\ref{eq_Umj_def} is given by
\begin{equation}
\bar{U}_{m,j}(\alpha) =  F_n\left[-\left(\frac{GM\mu}{R}\right) [b_{1/2}^{(m)}(\alpha) - q \delta_{(|m|,1)}\alpha] \right].
\label{eq_Umj_ma}
\end{equation}
Comparing Eqs.~\ref{eq_U} and \ref{eq_disturb_pot_ma},
from the unicity of the Fourier expansion, and 
from the linearity of the operators $F_n$,
we finally obtain for a generic potential as given by Eq.~\ref{eq_U}
\begin{equation}
\bar{U}_{m,j} (\alpha)= 2 F_n[U_m(\alpha)].
\label{eq_Umj_gen}
\end{equation}

This is the central equation of this Section, 
as it gives the amplitude $\bar{U}_{m,j} (\alpha)$ of any $m/(m-j)$ resonance term,  
whether internal or external, and whether direct or indirect in nature, 
and for any potential of the form of Eq.~\ref{eq_U},
\textit{i.e.} satisfying the conditions $(i)$-$(iii)$ at the start of Section~\ref{sec_prelem}.
As a word of caution, note that if Eq.~\ref{eq_U0} is used instead of Eq.~\ref{eq_U}, 
then we must use $\bar{U}_{m,j} (\alpha)= F_n[U_m(\alpha)]$ instead of Eq.~\ref{eq_Umj_gen}.

The operators $F_n$ for resonances of order 1, 2, 3 and 4 are listed in Table~\ref{tab_reson_terms}.

\begin{table}[!t]
\caption{Resonant terms $\bar{U}_{m,j}(\alpha)$ (Eq.~\ref{eq_Umj_def})}
\label{tab_reson_terms}
\begin{tabular}{ccl}
\hline
Order $j$ & $\bar{U}_{m,j}$ & Operators $F_n$ (Eq.~\ref{eq_Umj_gen}) \\
\hline
1 & $2 e     F_{27}[U_m(\alpha)] \cos(\phi_{m,1})$    & 
$F_{27}= (1/2)[-2m - \alpha D]$ (Lindblad resonances) \\
2 & $2 e^2 F_{45}[U_m(\alpha)] \cos(2\phi_{m,2})$  & 
$F_{45}= (1/8)[-5m + 4m^2 + (-2 + 4m)\alpha D + \alpha^2 D^2]$ \\
3 & $2 e^3 F_{82}[U_m(\alpha)] \cos(3\phi_{m,3})$  &
$F_{82}= (1/48)[-26m + 30m^2 - 8m^3 + (-9 + 27m - 12m^2)\alpha D$ \\
   &                                                                             & $+ (6 - 6m)\alpha^2D^2 - \alpha^3D^3]$ \\
4 & $2 e^4 F_{90}[U_m(\alpha)] \cos(4\phi_{m,4})$  & 
$F_{90}=  (1/384)[-206m + 283m^2 -120m^3 +16m^4$ \\
   &                                                                             & $+ (-64 + 236m -168m^2 + 32m^3)\alpha D$ \\
 & & 
$+ (48 - 78m + 24m^2)\alpha^2D^2 + (-12 + 8m)\alpha^3D^3 + \alpha^4D^4]$ \\
\hline
\end{tabular}
\tablecomments{%
The operators $F_n$ are found in \cite{mur00} or \cite{ell00}. 
The resonant argument $\phi_{m,j}$ is given in Eq.~\ref{eq_phi_mj} and 
$D$ is the radial derivative operator, $D=d/d\alpha$.
When applied to indirect terms, $\alpha^p D^p=  \alpha \delta_{(p,1)}$ (Eq.~\ref{eq_aD}).
In the case of a homogeneous triaxial ellipsoid, 
the operator $\alpha^pD^p$ reduces to a multiplicative factor
$\alpha^pD^p= (-1)^p(|m|+1)...(|m|+p)$ (Eq.~\ref{eq_apDp}).}
\end{table}

\subsection{Triaxial homogeneous ellipsoid}
\label{subsec_homo_ellipsoid}

We now consider the example of a triaxial homogeneous ellipsoid with semi-axes $A>B>C$, rotating around its minor axis $C$,
see details in \cite{sic19,sic20}.
The reference radius $R$ of the ellipsoid is defined by 
\begin{equation}
\frac{3}{R^2}= \frac{1}{A^2} + \frac{1}{B^2} + \frac{1}{C^2},
\label{eq_r}
\end{equation}
and its elongation and oblateness are measured by the dimensionless parameters $\epsilon$ and $f$  
\begin{equation}
\epsilon = \frac{A^2-B^2}{2R^2} {\rm~~and~~} 
f = \frac{A^2+B^2-2C^2}{4R^2}.
\label{eq_epsilon_f}
\end{equation}

Because of the symmetry of the body, its potential is invariant under a $\pi$-rotation, 
so that only even values of $m$ appear in the Fourier expansion in Eq.~\ref{eq_U}, 
thus eliminating the indirect part of the potential.
Posing $m=2p$, the resonance condition \ref{eq_ratio_n_omega_approx} now reads 
\begin{equation}
\frac{n}{\Omega} \sim \frac{2p}{2p-j},
\label{eq_ratio_n_omega_approx_ell}
\end{equation}
which eliminates (among others) every other Lindblad resonances, keeping only those with $m$ even.
For instance, the 4/3 resonance ($m=4, j=1$) survives as a Lindblad resonance, 
while the 5/4 resonance vanishes, 
leaving its place to the second-order resonance 10/8 resonance ($m=10, j=2$).

At lower order in $\epsilon$ and $f$, $U(\textbf{r})$ is given by \cite{sic19,sic20}
\begin{equation}
U(\textbf{r}) = -\frac{GM}{R} 
\sum_{m=-\infty}^{+\infty} 
\left( \frac{R}{r} \right)^{|m|+1} 
\epsilon^{|m/2|} S_{|m/2|}
\cos\left(m \theta \right) ~~(m{\rm~even}),
\label{eq_pot_ell}
\end{equation}
where $S_{|p|}$ is recursively defined by
\begin{equation}
S_{|p|+1} = 2
\frac{(|p|+1/4)(|p|+3/4)}{(|p|+1)(|p|+5/2)}
\times S_{|p|}
{\rm~~with~~}
S_0 = 1.
\label{eq_Sp}
\end{equation}
By comparing Eqs.~\ref{eq_U} and \ref{eq_pot_ell}, we obtain
%
\begin{equation}
U_m(\alpha)=  -\left( \frac{GM}{R} \right)
\frac{\epsilon^{|m/2|} S_{|m/2|}}{\alpha^{|m|+1}} ~~(m{\rm~even}),
\label{eq_Um_ell}
\end{equation}
where again $\alpha$ is given by Eq.~\ref{eq_alpha}.
Due to the form of $U_m(\alpha)$, 
a power of $\alpha$, 
the differential operator $\alpha^p D^p$ reduces here to a mere multiplicative factor
\begin{equation}
\alpha^p D^p = (-1)^p (|m|+1)...(|m|+p) ~~(m{\rm~even}), 
\label{eq_apDp}
\end{equation}
so that the operators $F_n$ in Table~\ref{tab_reson_terms}
are multiplicative factors that are polynomial functions of $m$ and $|m|$. From Eq.~\ref{eq_Umj_gen},
\begin{equation}
\bar{U}_{m,j}(\alpha)=
-\left(\frac{GM}{R} \right)
\epsilon^{|m/2|}
\left(\frac{2 S_{|m/2|}F_n}{\alpha^{|m|+1}}\right) 
e^j 
\cos(j\phi_{m,j}) ~~(m{\rm~even}),
\label{eq_Umj_ell}
\end{equation}
%
%

This is a convenient way to express $\bar{U}_{m,j}(\alpha)$ as the product of 
\begin{enumerate}
\item a potential term $-GM/R$ that globally scales the problem in terms of mass and length,
\item a dimensionless term $\epsilon^{|m/2|}$ that measures the departure of the body from axisymmetry
(akin to a mass anomaly),
\item a dimensionless factor $2 S_{|m/2|}F_n/\alpha^{|m|+1}$ that is intrinsic to the resonance,
{\it i.e.} to the azimuthal number $m$ (Table~\ref{tab_reson_terms}) and the order $j$, through the value of $\alpha$,
\item a term $e^j$ that defines the resonance order, and
\item a trigonometric term $\cos(j\phi_{m,j})$, where $\phi_{m,j}$ is defined by Eq.~\ref{eq_phi_mj}.
\end{enumerate}

In order to isolate what is intrinsic to the resonance geometry and to the non-axisymmetry of the body, 
I define the strength of a $m/(m-j)$ resonance as the dimensionless coefficient
\begin{equation}
{\cal S}_{m,j} =
\epsilon^{|m/2|}
\left(\frac{2 S_{|m/2|}F_n}{\alpha^{|m|+1}}\right) ~~(m{\rm~even}).
\label{eq_strength}
\end{equation}

The factors $F_n$ are given in Table~\ref{tab_reson_terms} and $S_{|m/2|}$ is defined in Eq.~\ref{eq_Sp}.
The factor $\alpha$ can be calculated from the condition $j\kappa(a)= m[n(a)-\Omega]$ and 
the expressions of $n(a)$ and $\kappa(a)$ as a function of $GM$, $\epsilon$ and $f$. 
To lowest order in $\epsilon$ and $f$, we have from Eq.~\ref{eq_n_kappa} and 
\cite{sic19,sic20}\footnote{If need be, higher order terms in $f$ and $\epsilon$ can be introduced 
in Eq.~\ref{eq_n_kappa_approx}, using the expansions of \cite{sic19,sic20}.}
\begin{equation}
\begin{array}{ccc}
\displaystyle
n^2(r) \sim \frac{GM}{a^3} \left[1 + \frac{3f}{5} \left(\frac{R}{a}\right)^2\right] &
{\rm and} & 
\displaystyle
\kappa^2(r) \sim \frac{GM}{a^3} \left[1 - \frac{3f}{5} \left(\frac{R}{a}\right)^2\right], \\
\end{array}
\label{eq_n_kappa_approx}
\end{equation}
and $\alpha= a/R$ can be numerically determined through an iterative process if $f$ is sufficiently small,
see \textit{e.g.} \citealt{ren06}.

As examples, the factors ${\cal S}_{m,j}e^j$ are listed in Table~\ref{tab_strength} 
in the cases of Chariklo and Haumea,
assumed to be homogeneous triaxial bodies. The following points can be noted:
\begin{enumerate}
\item
because of the term $\epsilon^{|m/2|}$, the resonance strength rapidly tends to zero as $m$ tends to infinity, 
{\it i.e.} as one approaches the corotation radius, see an example in Fig.~2 of \cite{sic19}.
This contrasts with the case of a perturbing satellite, for which ${\cal S}_{m,j}$ increases as
$m$ increases (for $j$ fixed), since the particles orbit closer and closer to the satellite, 
\item
some resonances are not replicated Table~\ref{tab_strength} (symbols ****) 
because only the lowest order in eccentricity has been considered for a given ratio $n/\Omega$.
For instance, the $m = -2, j = 2$ case, 
corresponding to the second-order $\epsilon e^2$-resonance ($n/\Omega = 2/4$),
is not considered in its fourth-order version $\propto \epsilon^2 e^4$ 
with $m = -4, j = 4$ ($n/\Omega = 4/8$).
\end{enumerate}

\begin{table}[!t]
\caption{Resonance strengths around homogenous ellipsoids}
\label{tab_strength}
\begin{tabular}{c|cccccccc}
\hline
\hline
                          & \multicolumn{8}{c}{Azimuthal number} \\
$m \rightarrow$ &-8 & -6 & -4 & -2 & 2 & 4 & 6 & 8 \\
\hline
\hline
Order $j \downarrow$ & \multicolumn{8}{c}{Chariklo\tablenotemark{a}} \\
\hline
1                       & $0.0105 \epsilon^4e$ 
                         & $0.0207 \epsilon^3e$ 
                         & $0.0439 \epsilon^2e$
                         & $0.102   \epsilon    e$
                         & inside
                         & $-0.0408 \epsilon^2e$
                         & $-0.0200 \epsilon^3e$
                         & $-0.0103 \epsilon^4e$
                         \\
                         &  210 [8/9]
                         &  214 [6/7]
                         &  223 [4/5]
                         & 250 [2/3]
                         & [2/1]
                         & 169 [4/3]
                         & 178 [6/5]
                         & 182 [8/7]
                         \\
\hline
2                       & **** 
                         & $0.0641 \epsilon^3 e^2$
                         & **** 
                         & $0.143   \epsilon e^2$
                         & apsidal
                         & **** 
                         & $0.0336   \epsilon^3 e^2$
                         & **** 
                         \\
                         & 223 [8/10]
                         & 232 [6/8]
                         & 250 [4/6]
                         & 300 [2/4]
                         & [2/0]
                         & [4/2]
                         & 160 [6/4]
                         & 169 [8/6]
                         \\
\hline
3                       & $0.125  \epsilon^4e^3$ 
                         & **** 
                         & $0.185 \epsilon^2e^3$
                         & $0.190 \epsilon e^3$
                         & inside
                         & inside
                         & **** 
                         & inside
                         \\
                         &  237 [8/11]
                         &  250 [6/9]
                         &  275 [4/7]
                         &  348 [2/5]
                         & [2/-1]
                         & [4/1]
                         & [6/3]
                         & [8/5]
                         \\
\hline
4                       & **** 
                         & $0.331 \epsilon^3e^4$
                         & **** 
                         & $0.251 \epsilon e^4$
                         & $0.00251 \epsilon e^4$
                         & **** 
                         & inside
                         & **** 
                         \\
                         &  250 [8/12]
                         &  267 [6/10]
                         &  300 [4/8]
                         &  392 [2/6]
                         & 196 [2/-2]
                         & [4/0]
                         & [6/2]
                         & [8/4]
                         \\
\hline
\hline
Order $j \downarrow$ & \multicolumn{8}{c}{Haumea\tablenotemark{b}} \\
\hline
\hline
1                       & $0.0163  \epsilon^4e$ 
                         & $0.0294  \epsilon^3e$ 
                         & $0.0570  \epsilon^2e$
                         & $0.121     \epsilon e$
                         & inside
                         & inside
                         & inside
                         & inside
                         \\
                         & 1238 [8/9]
                         & 1263 [6/7]
                         & 1313 [4/5]
                         & 1463 [2/3]
                         & [2/1]
                         & [4/3]
                         & [6/5]
                         & [8/7]
                         \\
\hline
2                       & **** 
                         & $0.0937  \epsilon^3e^2$
                         & **** 
                         & $0.171    \epsilon e^2$
                         & apsidal
                         & **** 
                         & inside
                         & **** 
                         \\
                         & 1313 [8/10]
                         & 1363 [6/8]
                         & 1463 [4/6]
                         & 1752 [2/4]
                         & [2/0]
                         & [4/2]
                         & [6/4]
                         & [8/6]
                         \\
\hline
3                       & $0.204  \epsilon^4e^3$ 
                         & **** 
                         & $0.248  \epsilon^2e^3$
                         & $0.229  \epsilon e^3$
                         & inside
                         & inside
                         & **** 
                         & inside
                         \\
                         & 1388 [8/11]
                         & 1463 [6/9]
                         & 1609 [4/7]
                         & 2025 [2/5]
                         & [2/-1]
                         & [4/1]
                         & [6/3]
                         & [8/5]
                         \\
\hline
4                       & **** 
                         & $0.500 \epsilon^3e^4$
                         & **** 
                         & $0.302 \epsilon e^4$
                         & $0.00286 \epsilon e^4$
                         & **** 
                         & inside
                         & **** 
                         \\
                         & 1463 [8/12]
                         & 1561 [6/10]
                         & 1752 [4/8]
                         & 2285 [2/6]
                         & 1164 [2/-2]
                         & [4/0]
                         & [6/2]
                         & [8/4]
                         \\
\hline
\hline
\end{tabular} \\
\tablenotetext{a}{%
Using 
$M= 6.3 \times 10^{18}$~kg, 
$T_{\rm rot} = 2\pi/\Omega = 7.004$~h, 
$A\times B \times C =  57 \times 139 \times 86$~km,
$R = 115$~km, 
$f= 0.20$, $\epsilon= 0.61$
\citep{lei17}.
}%
\tablenotetext{b}{%
Using
$M= 4.006 \times 10^{21}$~kg,
$T_{\rm rot} =  3.915341$~h, 
$A\times B \times C = 1161 \times 852 \times 513$~km,
$R = 712$~km, 
$f= 0.55$, $\epsilon= 0.76$
\citep{ort17}.
}%
\tablecomments{%
In each box, the factor ${\cal S}_{m,j}e^j$ is calculated for the specified values of $m$ and $j$,
using Eq.~\ref{eq_strength}.
Below each factor are the corresponding resonant radius (km) and the ratio $[n/\Omega]$. 
The term ``inside" means that the resonance formally occurs inside the physical volume of the body,
and is thus unphysical.
Note that the apsidal resonances also occur formally inside the body, and are not treated here.
The **** symbols indicate resonances that are already listed elsewhere in the Table
under a lower order version, see text.}%
\end{table}

\section{Concluding remarks}
\label{sec_concl}

In this paper, I have investigated the structure of the $j\kappa= m(n-\Omega)$ sectoral resonances
in the equatorial plane of a non-axisymmetric object rotating at rate $\Omega$. 
%
%
The cases $j=0$ (corotation) and $m=j$ (apsidal) are not studied here.
Fig.~\ref{fig_taxonomy} summarizes the general taxonomy for those resonances and 
Fig.~\ref{fig_braids_self_crossing} illustrates some of the results on the structure of resonant orbits.

The kinematic structure of a resonant orbit associated with $(m,j)$ is entirely
encapsulated in the couple $(m',j')$, the irreducible (relatively prime) version of $(m,j)$.
Thus, the kinematic structure of the orbit only depends on $n/\Omega \sim m/(m-1)= m'/(m'-j')$, 
\textit{i.e.} on the resonance location, and is independent of the nature of the potential.
More precisely, the resonant orbit has $j'$ braids, $|m'|$ identical sectors and $|m'|(j'-1)$ self-crossing points.

The existence of a resonance, and therefore its order $j$ for a given $n/\Omega$ ratio, 
depends on the symmetry of the potential. 
In particular, a potential that is invariant under a $2\pi/k$-rotation creates only 
resonances of the form $kp/(kp-j)$.
This is why, for instance, the second-order 1/3 resonance around a spherical body with a mass anomaly,
which has $m=-1$, $j=2$, $k=1$, 
is replaced by the fourth-order resonance 2/6 around a homogeneous ellipsoid, 
which has $m=-2$, $j=4$, $k=2$.

Resonances that have opposite $m$ and same $j$ have periodic orbits that possess the same kinematic structure
and the same order, \textit{i.e.} the same dynamical behavior. 
Here, they are called \textit{true twin} resonances.
Resonances with opposite $m'$ and same $j'$, but different $|m|$ and $j$ are called
\textit{false twin} resonances, because they correspond to the same kinematic, 
but to different dynamical behaviors.

A retrograde resonance ($n/\Omega < 0$) is always of higher order than 
the corresponding prograde resonance occurring at the same radius, but with $n/\Omega > 0$. 
This shows that there are no retrograde Lindblad ($j=1$) resonances.

The resonance strengths can be calculated using a unique operator 
(for a given couple $(m,j)$)
that acts on the direct and indirect parts of the potential, 
and that is valid for inner, outer, prograde and retrograde resonances, see Eq.~\ref{eq_Umj_gen}.
These operators are in fact the classical operators $F_n$ used for satellite perturbations.
In the case of a homogeneous triaxial ellipsoid, they reduce to mere multiplicative factors
(Table~\ref{tab_reson_terms}) that are easily implemented in numerical schemes.
Examples are given in Table~\ref{tab_strength} for
Chariklo and Haumea, assumed to be homogeneous triaxial ellipsoids.

This study is intended to be general enough to serve in a broad range of contexts.
For instance, the results can easily be generalized in cases where the central body has several equatorial mass anomalies.
Then, it is enough to split the potential in elementary, single-anomaly potentials, 
and accounting for the fact that those potentials are out of phase. 

As mentioned earlier, Eq.~\ref{eq_streamline} can be seen as describing a streamline of particles in a collisional disk. 
Then difficulties arise because self-crossing cause singularities in the hydrodynamical equations 
that describe the flow of particles near the resonance.
Moreover, and except for the Lindblad resonances, 
these equations involve non-linear perturbations because they are of order $j>1$ in eccentricity, 
a further source of complications.

However, not having the appropriate hydrodynamical equations does mean that those resonances have no effects on the disk.
In that context, it would be interesting to use works already done on granular flows or kinetic theories
to describe neighbor-streamline crossings in waves excited by Lindblad resonances, see \textit{e.g.} \cite{bor85,shu85}.
Another approach is to rely on collisional codes that include a realistic description of particulate collisions in rings.
This can be relevant to Chariklo's and Haumea's rings, 
as both ring systems are found to orbit near the second-order 1/3 (or fourth-order 2/6) resonance 
with their host body \citep{ort17,sic20},
a subject of future works. 

\acknowledgments

The work leading to these results has received funding from the 
European Research Council under the European Community's H2020
2014-2020 ERC Grant Agreement n$^\circ$ 669416 ``Lucky Star".
I thank Fran\c{c}oise Combes, Renu Malhotra and Scott Tremaine
for discussions when preparing this paper.

\end{document}